\documentclass[a4paper,fleqn]{article}
\usepackage{amsmath,amsfonts}
\begin{document}

\title{\textbf{Integrability study of a four-dimensional eighth-order nonlinear wave equation}}

\author{\textsc{Sergei Sakovich}\bigskip \\
\small Institute of Physics, National Academy of Sciences of Belarus \\
\small sergsako@gmail.com}

\date{}

\maketitle

\begin{abstract}
We study the integrability of the four-dimensional eighth-order nonlinear wave equation of Kac and Wakimoto, associated with the exceptional affine Lie algebra ${\mathfrak e}_6^{(1)}$. Using the Painlev\'{e} analysis for partial differential equations, we show that this equation must be non-integrable in the Lax sense but very likely it possesses a lower-order integrable reduction.
\end{abstract}

\section{Introduction}

Most of the known integrable nonlinear wave equations, or soliton equations, are two-dimensional ones. In the literature, there is an increasing interest in higher-dimensional integrable equations. Some remarkable solitary wave solutions were found in higher dimensions, such as the line solitons, lumps, dromions, etc. There is a strong demand for higher-dimensional integrable models in physics, especially in nonlinear optics, field theory, hydrodynamics, and plasma physics. The development of new methods to analyze and solve higher-dimensional nonlinear equations can stimulate many branches of pure and applied mathematics. However, only several three-dimensional integrable nonlinear equations are known at present, and even less is known about the integrability in dimension four. For these reasons, every new higher-dimensional nonlinear equation, reported in the literature to be integrable in some sense, deserves a comprehensive investigation.

In the present paper, we study the integrability of the Kac--Wakimoto four-dimensional eighth-order nonlinear wave equation associated with the exceptional affine Lie algebra ${\mathfrak e}_6^{(1)}$, which has the following Hirota bilinear form \cite{KW}:
\begin{equation}
\left( D_x^8 - 280 \sqrt{6} D_x^3 D_y + 210 D_z^2 - 240 \sqrt{2} D_x D_t \right) \tau \cdot \tau = 0 , \label{e1}
\end{equation}
where $\tau = \tau (x,y,z,t)$, and the Hirota differentiation operators are defined by $D_x^n \tau \cdot \tau = ( \partial_x - \partial_{x'} )^n \tau(x,y,z,t) \tau(x',y,z,t) |_{x' = x}$ and similar relations for $y,z,t$. Due to construction, this nonlinear equation \eqref{e1} possesses multi-soliton solutions containing arbitrarily many free parameters \cite{KW}. Recently, some exact one-soliton and two-soliton solutions of \eqref{e1} were found and studied by Dodd \cite{D}. Note that the Kac--Wakimoto equation \eqref{e1} was called integrable in \cite{D}. The existence of a multi-soliton solution, however, not necessarily implies the existence of a good Lax representation for a studied nonlinear wave equation. No Lax pair has been found for the Kac--Wakimoto equation \eqref{e1} as yet. Therefore it is reasonable to investigate the integrability of \eqref{e1} by a different method, and we do this by means of the Painlev\'{e} analysis for partial differential equations \cite{WTC,T,H}, like we did in \cite{S1} for some other nonlinear equations possessing multi-soliton solutions.

\section{The Painlev\'{e} analysis}

We consider the Painlev\'{e} analysis as a reliable and easy-to-use tool to test the integrability of nonlinear wave equations, especially convenient (in comparison with other integrability tests) for high-dimensional, high-order, multi-component and non-evolutionary equations \cite{S1,S2,S3,S4,KS1,KSY,KS2,S5,S6,S7,S8}. The reliability of the Painlev\'{e} test has been empirically verified in many integrability studies of multi-parameter nonlinear equations, including the fifth-order KdV-type equation \cite{HO}, the coupled KdV equations \cite{K,S9} (see further details in \cite{S10,KKS1,S11}), the symmetrically coupled higher-order nonlinear Schr\"{o}dinger equations \cite{ST1} (see also \cite{ST2,ST3}), the generalized Ito system \cite{KKS2}, the sixth-order bidirectional wave equation \cite{KKSST}, and the seventh-order KdV-type equation \cite{X}.

The transformation
\begin{equation}
u = 2 \frac{\tau_x}{\tau} \label{e2}
\end{equation}
brings the Kac--Wakimoto equation \eqref{e1} into the following form, appropriate to start the Painlev\'{e} analysis:
\begin{gather}
u_{8x} + 28 u_x u_{6x} + 28 u_{2x} u_{5x} + 70 u_{3x} u_{4x} \notag \\
+ 210 u_x^2 u_{4x} + 420 u_x u_{2x} u_{3x} + 420 u_x^3 u_{2x} \notag \\
+ a ( u_{3x,y} + 3 u_y u_{2x} + 3 u_x u_{x,y} ) + b u_{2z} + c u_{x,t} = 0 , \label{e3}
\end{gather}
where derivatives are denoted as $u_{3x,y} = \partial_x^3 \partial_y u$ and the like, and
\begin{equation}
a = - 280 \sqrt{6} , \qquad b = 210 , \qquad c =  - 240 \sqrt{2} . \label{e4}
\end{equation}
In what follows, we do not use these relations \eqref{e4}, that is, we study the nonlinear equation \eqref{e3} with $a,b,c$ being arbitrary parameters. This is convenient for several reasons. The values of $a,b,c$ in \eqref{e3} can be changed by scale transformations of $y,z,t$, respectively. As the result, there is only a finite set of essentially different values: $a = 1,0$, $b = 1,0$ (or $b = 1,0,-1$ if complex-valued transformations are not allowed), and $c = 1,0$. The original case \eqref{e4} is equivalent to the case $a=b=c=1$. The cases of \eqref{e3} with $a=0$, $b=0$ or $c=0$ are also interesting, because they correspond via \eqref{e2} to lower-dimensional reductions of the Kac--Wakimoto equation \eqref{e1}. For example, the case of \eqref{e3} with $b=0$ covers the Kac--Wakimoto equation associated with the exceptional affine Lie algebra ${\mathfrak d}_4^{(3)}$ \cite{KW}, which is the $z$-independent reduction of \eqref{e1}. Since the terms of \eqref{e3} with the coefficients $a,b,c$ are non-dominant terms during the Painlev\'{e} analysis, the values of $a,b,c$ play no role up to the last step of the analysis, where the compatibility conditions at the resonances are checked, and this makes possible to study all the cases of essentially different values of $a,b,c$ simultaneously.

The nonlinear wave equation \eqref{e3} is a normal system of order eight and dimension four, therefore its general solution must contain eight arbitrary functions of three variables. A hypersurface $\phi(x,y,z,t) = 0$ is non-characteristic for this equation if $\phi_x \neq 0$, and we set $\phi_x = 1$ without loss of generality, that is
\begin{equation}
\phi = x + \psi(y,z,t) \label{e5}
\end{equation}
with an arbitrary function $\psi$. Looking for a singular behavior of solutions $u$ of the nonlinear equation \eqref{e3} near a hypersurface $\phi = 0$, in the form
\begin{equation}
u = u_0 (y,z,t) \phi^{\alpha} + \dotsb , \label{e6}
\end{equation}
we find that there is only one admissible value of the leading exponent $\alpha$, and that three different values of the coefficient $u_0$ correspond to that value of $\alpha$:
\begin{equation}
\alpha = -1 , \qquad u_0 = 2,4,6 . \label{e7}
\end{equation}
Consequently, there are three different branches of a singular (pole-like) behavior of solutions of \eqref{e3} near an arbitrary non-characteristic hypersurface.

Substituting the expansion
\begin{equation}
u = u_0 \phi^{-1} + \dotsb + u_r (y,z,t) \phi^{r-1} + \dotsb \label{e8}
\end{equation}
to the nonlinear equation \eqref{e3}, and collecting terms with $\phi^{r-9}$, we find the following positions $r$ of the resonances, where arbitrary functions of $y,z,t$ can enter the singular expansion of a solution, separately for each of the three branches:
\begin{gather}
u_0 = 2 , \qquad r = -1, 1, 2, 3, 4, 5, 8, 14, \label{e9} \\
u_0 = 4 , \qquad r = -2, -1, 1, 2, 3, 8, \frac{1}{2} \left( 25 \pm \sqrt{65} \right) , \label{e10} \\
u_0 = 6 , \qquad r = -3, -2, -1, 1, 8, 10, \frac{1}{2} \left( 23 \pm \sqrt{193} \right) . \label{e11}
\end{gather}
Taking into account that $r = -1$ corresponds to the arbitrariness of the function $\psi$ in \eqref{e5}, we conclude that the expansion \eqref{e8} with $u_0 = 2$ may represent the general solution of \eqref{e3}. In this generic branch \eqref{e9}, all the resonances lie in integer positions, what is appropriate for the Painlev\'{e} property. This is, however, not the case for the non-generic branches \eqref{e10} and \eqref{e11}, where the pairs of resonances lie in non-integer, irrational positions, what indicates a kind of infinite branching of some special solutions of \eqref{e3}. Right at this step of the Painlev\'{e} analysis, we can conclude that the nonlinear equation \eqref{e3} does not possess the Painlev\'{e} property due to inappropriate positions of resonances. Let us note that this result does not depend on the values of $a,b,c$, because the terms of \eqref{e3} with the coefficients $a,b,c$ are non-dominant terms during the Painlev\'{e} analysis and have no influence on positions of resonances therefore. In other words, the four-dimensional eighth-order nonlinear equation \eqref{e1} fails the Painlev\'{e} test in absolutely the same way as does the one-dimensional (ordinary differential) eighth-order nonlinear equation $D_x^8 \; \tau \cdot \tau = 0$ which is the $(y,z,t)$-independent reduction of \eqref{e1}.

The non-integer positions of resonances, however, cannot serve as an ultimate indication that the nonlinear equation \eqref{e3} is non-integrable. Some Lax integrable nonlinear wave equations have resonances in rational positions \cite{S12} (this is the so-called weak Painlev\'{e} property), some integrable by quadratures ordinary differential equations show even irrational positions of resonances \cite{S13} (though we do not know such partial differential equations). For this reason, let us continue the analysis in the generic branch \eqref{e9}, in order to find a stronger indication of non-integrability, the strongest possible in the framework of the Painlev\'{e} analysis.

Substituting the expansion
\begin{equation}
u = \sum_{i=0}^{\infty} u_i (y,z,t) \phi^{i-1} \label{e12}
\end{equation}
to the nonlinear equation \eqref{e3}, and collecting terms with $\phi^{n-9}$, $n = 0,1,2, \dotsc$, we obtain the recursion relations for the coefficients $u_n$ of the expansion, which either determine the function $u_n (y,z,t)$ in terms of the functions $u_0 , \dotsc , u_{n-1} , \psi$ and their derivatives, if $n$ is not a resonance position, or determine a compatibility condition for the functions $u_0 , \dotsc , u_{n-1} , \psi$ and their derivatives, if $n$ is a resonance position. One should definitely use a computer algebra system to reproduce the subsequent calculations (we used Mathematica 5.2 \cite{W} to do them), and we omit unnecessary cumbersome expressions therefore. We set $u_0 = 2$. This satisfies the recursion relations at $n=0$ and means that we study the generic branch with positions of resonances given by \eqref{e9}. At $n = 1,2,3,4,5$, we have the resonances, the compatibility conditions turn out to be satisfied identically, and $u_1 , u_2 , u_3 , u_4 , u_5$ remain arbitrary functions of $y,z,t$. At $n = 6,7$, the recursion relations give expressions for $u_6 , u_7$, respectively. At the resonance $n = 8$, the arbitrary function $u_8 (y,z,t)$ appear, and the compatibility condition is satisfied identically. Next, at $n = 9,10,11,12,13$, the recursion relations give expressions for $u_9 , u_{10} , u_{11} , u_{12} , u_{13}$, respectively. Finally, at the resonance $n = 14$, where the arbitrary function $u_{14} (y,z,t)$ enters the expansion \eqref{e12}, the compatibility condition is not satisfied identically but has the form
\begin{equation}
u_8 = \frac{1}{u_3^2} P[u_1 , u_2 , u_3 , u_4 , u_5 ; a , b , c] , \label{e13}
\end{equation}
where $P$ denotes a complicated polynomial expression involving the functions $u_1 , u_2 , u_3 , u_4 , u_5$, their derivatives, and the parameters $a,b,c$. The explicit form of $P$ is not required to conclude that the nonlinear equation \eqref{e3} is non-integrable. The fact, itself, that the recursion relations are not compatible at the resonance $n = 14$ indicates that the Laurent-type expansion \eqref{e12} does not represent the general solution of \eqref{e3}, and that we should modify \eqref{e12} by adding logarithmic terms, starting from the term proportional to $\phi^{13} \log \phi$. No examples of integrable equations with non-dominant logarithmic branching of solutions are known. This type of singularities is generally considered as an ultimate indicator of non-integrability of nonlinear differential equations. Let us also note that the condition \eqref{e13} is not an identity for any values of the parameters $a,b,c$ in \eqref{e3}, zero or non-zero. Therefore our conclusion on non-integrability of the four-dimensional eighth-order nonlinear equation \eqref{e1} is valid for the $y$-, $z$- and $t$-independent reductions of this equation as well, including the Kac--Wakimoto equation associated with the exceptional affine Lie algebra ${\mathfrak d}_4^{(3)}$ \cite{KW}.

\section{Conclusion}

Taking into account the obtained results of the Painlev\'{e} analysis, we believe that the Kac--Wakimoto equation \eqref{e1} cannot possess any good Lax representation. We believe, however, that this non-integrable four-dimensional eighth-order nonlinear equation can possess an integrable four-dimensional lower-order reduction. The existence of such a reduction could explain why the Kac--Wakimoto equation \eqref{e1}, being (most probably) non-integrable itself, possesses multi-soliton solutions in dimension four. An integrable four-dimensional lower-order reduction of the nonlinear equation \eqref{e3} may exist because the nontrivial compatibility condition \eqref{e13} fixes only one of the eight arbitrary functions involved in the Laurent-type expansion \eqref{e9} and does not reduce the number of independent variables the remaining seven arbitrary functions depend on. We believe that it is possible to find this integrable reduction by means of the truncated singular expansion technique \cite{WTC,T,H}, and the work in this direction is in progress.

\section*{Acknowledgments}

The author is deeply grateful to Prof. V.G. Kac for his kind help and interest, and to the Max Planck Institute for Mathematics, where a part of this work was done.


\begin{thebibliography}{99}

\bibitem{KW} V.G. Kac, M. Wakimoto, Exceptional hierarchies of soliton equations, Proc. Sympos. Pure Math. 49 (1989) 191--237.

\bibitem{D} R.K. Dodd, An integrable equation associated with ${\mathfrak e}_6^{(1)}$, Phys. Lett. A 372 (2008) 6887--6889.

\bibitem{WTC} J. Weiss, M. Tabor, G. Carnevale, The Painlev\'{e} property  for partial differential equations, J. Math. Phys. 24 (1983) 522--526.

\bibitem{T} M. Tabor, Chaos and Integrability in Nonlinear Dynamics: An Introduction, Wiley, New York, 1989.

\bibitem{H} A.N.W. Hone, Painlev\'{e} tests, singularity structure and integrability, In: A.V. Mikhailov (Ed.), Integrability, Lect. Notes in Phys. 767, Springer, Berlin, 2009, 245--277; arXiv:nlin.SI/0502017.

\bibitem{S1} S.Yu. Sakovich, Painlev\'{e} analysis of new soliton equations by Hu, J. Phys. A 27 (1994) L503--L505.

\bibitem{S2} S.Yu. Sakovich, Painlev\'{e} analysis and B\"{a}cklund transformations of Dok\-torov--Vlasov equations, J. Phys. A 27 (1994) L33--L38.

\bibitem{S3} S.Yu. Sakovich, Painlev\'{e} analysis of a higher-order nonlinear Schr\"{o}dinger equation, J. Phys. Soc. Jpn. 66 (1997) 2527--2529.

\bibitem{S4} S.Yu. Sakovich, On integrability of a $(2+1)$-dimensional perturbed KdV equation, J. Nonlinear Math. Phys. 5 (1998) 230--233; arXiv:solv-int/9805012.

\bibitem{KS1} A. Karasu-Kalkanl\i, S.Yu. Sakovich, B\"{a}cklund transformation and special solutions for the Drinfeld--Sokolov--Satsuma--Hirota system of coupled equations, J. Phys. A 34 (2001) 7355--7358; arXiv:nlin/0102001.

\bibitem{KSY} A. Karasu-Kalkanl\i, S.Yu. Sakovich, \'{I}. Yurdu\c{s}en, Integrability of Kersten--Krasil'shchik coupled KdV--mKdV equations: singularity analysis and Lax pair, J. Math. Phys. 44 (2003) 1703--1708; arXiv:nlin/0206046.

\bibitem{KS2} A. Karasu-Kalkanl\i, S. Sakovich, Singularity analysis of a spherical Ka\-domtsev--Petviashvili equation, J. Phys. Soc. Jpn. 74 (2005) 505--507; arXiv:nlin/0404037.

\bibitem{S5} S. Sakovich, Enlarged spectral problems and nonintegrability, Phys. Lett. A 345 (2005) 63--68; arXiv:nlin/0504037.

\bibitem{S6} S. Sakovich, Integrability of the vector short pulse equation, J. Phys. Soc. Jpn. 77 (2008) 123001; arXiv:0801.3179.

\bibitem{S7} S. Sakovich, Singularity analysis and integrability of a Burgers-type system of Foursov, SIGMA 7 (2011) 002; arXiv:1010.5709.

\bibitem{S8} S. Sakovich, On two aspects of the Painlev\'{e} analysis, Int. J. Analysis 2013 (2013) 172813; arXiv:solv-int/9909027.

\bibitem{HO} H. Harada, S. Oishi, A new approach to completely integrable partial differential equations by means of the singularity analysis, J. Phys. Soc. Jpn. 54 (1985) 51--56.

\bibitem{K} A. Karasu-Kalkanl\i, Painlev\'{e} classification of coupled Korteweg--de~Vries systems, J. Math. Phys. 38 (1997) 3616--3622.

\bibitem{S9} S.Yu. Sakovich, Coupled KdV equations of Hirota--Satsuma type, J. Nonlinear Math. Phys. 6 (1999) 255--262; arXiv:solv-int/9901005.

\bibitem{S10} S.Yu. Sakovich, Addendum to: Coupled KdV equations of Hirota--Satsuma type, J. Nonlinear Math. Phys. 8 (2001) 311--312; arXiv:nlin/0104072.

\bibitem{KKS1} A. Karasu-Kalkanl\i, A. Karasu, S.Yu. Sakovich, A strange recursion operator for a new integrable system of coupled Korteweg--de~Vries equations, Acta Appl. Math. 83 (2004) 85--94; arXiv:nlin/0203036.

\bibitem{S11} S. Sakovich, A note on the Painlev\'{e} property of coupled KdV equations, Int. J. Part. Diff. Eqns. 2014 (2014) 125821; arXiv:nlin/0402004.

\bibitem{ST1} S.Yu. Sakovich, T. Tsuchida, Symmetrically coupled higher-order nonlinear Schr\"{o}dinger equations: singularity analysis and integrability, J. Phys. A 33 (2000) 7217--7226; arXiv:nlin/0006004.

\bibitem{ST2} S.Yu. Sakovich, T. Tsuchida, Coupled higher-order nonlinear Schr\"{o}dinger equations: a new integrable case via the singularity analysis, arXiv: nlin/0002023.

\bibitem{ST3} S.Yu. Sakovich, T. Tsuchida, A new integrable system of symmetrically coupled derivative nonlinear Schr\"{o}dinger equations via the singularity analysis, arXiv:nlin/0004025.

\bibitem{KKS2} A. Karasu-Kalkanl\i, A. Karasu, S.Yu. Sakovich, Integrability of a generalized Ito system: the Painlev\'{e} test, J. Phys. Soc. Jpn. 70 (2001) 1165--1166; arXiv:nlin/0102030.

\bibitem{KKSST} A. Karasu-Kalkanl\i, A. Karasu, A. Sakovich, S. Sakovich, R. Turhan, A new integrable generalization of the Korteweg--de~Vries equation, J. Math. Phys. 49 (2008) 073516; arXiv:0708.3247.

\bibitem{X} G.Q. Xu, The integrability for a generalized seventh-order KdV equation: Painlev\'{e} property, soliton solutions, Lax pairs and conservation laws, Phys. Scr. 89 (2014) 125201.

\bibitem{S12} S.Yu. Sakovich, The Painlev\'{e} property transformed, J. Phys. A 25 (1992) L833--L836.

\bibitem{S13} S.Yu. Sakovich, On integrability of differential constraints arising from the singularity analysis, J. Nonlinear Math. Phys. 9 (2002) 21--25; arXiv: nlin/0004037.

\bibitem{W} S. Wolfram, The Mathematica Book, 5th ed., Wolfram Media, Champaign, 2003.

\end{thebibliography}
\end{document}